\documentclass[12pt]{iopart}
\begin{document}
%
%


%
%
\newcommand{\Df}{\phi_0}
\newcommand{\Dc}{\Delta_C}
\newcommand{\Ds}{\Delta_S}
\newcommand{\Dcp}{\Dc^{\prime}}
\newcommand{\Dsp}{\Ds^{\prime}}
\newcommand{\Cp}{C^{\prime}}
\newcommand{\Sp}{S^{\prime}}
\newcommand{\CM}{C_{(-)}}
\newcommand{\CP}{C_{(+)}}
\newcommand{\SM}{S_{(-)}}
\newcommand{\SP}{S_{(+)}}
\newcommand{\CMp}{C^{\prime}_{(-)}}
\newcommand{\CPp}{C^{\prime}_{(+)}}
\newcommand{\SMp}{S^{\prime}_{(-)}}
\newcommand{\SPp}{S^{\prime}_{(+)}}
\newcommand{\qe}{q_e}
\newcommand{\Qop}{\hat{Q} }
\newcommand{\Qdag}{\Qop^{\dag}}
\title{Discrete-charge quantum circuits
in semiclassical approach}
\author{C. A. Utreras-D\'{\i}az\dag\ and J. C. Flores\ddag}
\address{\dag\ Instituto de F\'\i sica, Facultad de Ciencias,
Universidad Austral de Chile, Casilla 567, Valdivia, Chile}
\address{\ddag\ Departamento de F\'{\i}sica, Facultad de Ciencias,
Universidad de Tarapac\'a, Casilla 7-D Arica, Chile}

\date{March 10, 2006}
\ead{cutreras@uach.cl}
\ead{cflores@uta.cl}

\begin{abstract}
We discuss a new approach to describe mesoscopic systems, based on the ideas of 
quantum electrical circuits with charge discreteness. This approach has allowed 
us to propose a simple alternative descriptions of some mesoscopic systems, with 
interesting results for some mesoscopic systems. In his work, we show that the application
of the Bohr-Sommerfeld quantization rules to the Quantum $LC$ circuit with discrete
charge allows us to easily reproduce previous results.
\end{abstract}
\maketitle

\section{Introduction}

In earlier work (see Li and Chen~\cite{LI-CHEN,YOU-LI}, Flores~\cite{FLORES}, Flores and 
Utreras~\cite{FLORES-UTRERAS,FLORES-UTRERAS2}, Utreras and Flores~\cite{UTRERAS-FLORES}), 
a theory of the quantum electrical circuit has been put forward. In this theory, the discrete 
nature of the electrical charge has been included by considering charge operators having discrete 
spectrum, and based on a simplified treatment of such systems as quantum $LC$ circuits, that is, 
electrical systems described by two fenomelogical parameters: an inductance $L$, and a capacitance $C$. 
It has been argued that this theory may be used to describe the behaviour of mesoscopic 
systems, such as mesoscopic rings, quantum tranmission lines, and the open electron 
resonator~\cite{DUNCAN}. As it is known, when the transport dimension becomes comparable with the charge carrier 
coherence length, one must take into account not only the quantum mechanical properties of the electron 
system, but also the discrete nature of electric charge, which leads the concept of quantum $LC$ circuit with 
discrete charge. We believe that the theory has the potential to help us understand many more phenomena relating to the 
systems mentioned above, if used correctly, and hence to provide a simplified complementary view of the 
behaviour of mesoscopic systems in general. 

In this article, we consider the quantum electrical system using the semiclassical approach, which allows 
us to describe the system in an approximate way, of interest here is the relationship that 
existes between our system and the simple pendulum.

\section{Summary of Quantum Circuits}

In their pioneering article, Li and Chen ~\cite{LI-CHEN,YOU-LI} consider a quantum $LC$ circuit, 
described by a wavefunction $\Psi(q)$ in the (continuum) charge  ($q$) representation, 
\begin{equation}
\hat H \Psi (q) = -\frac{\hbar^2}{2 L} \frac{d^2}{d q^2} \Psi(q) + V(q) \Psi(q).
\end{equation}
In this description, the charge ($\hat q$ ) and flux ($\hat \Phi$) operators, satisfy the
usual commutation rules, $[ \hat q, \hat \Phi ] = i \hbar$, and may be represented by the operators
$\hat{ q} =  q $, and $\hat{ \Phi}  =  - i \hbar d/dq $. The charge operator posseses continuum 
eigenvalues ($q_0$), with delta function eigenstates, just as the single particle quantum mechanical operators
$\hat q$ and $\hat p$.

\begin{equation}
\hat{ q} \Psi_{q_0} (q) = q_0 \Psi_{q_0} (q) = q_0 \delta(q-q_0) \nonumber .
\end{equation}


Li and Chen introduce discrete charge eigenstates, whith eigenvalues equal to multiples of the 
quantum of charge (the electron charge, $q_e$), $q_n = n q_e$ (integer $n$), 

\begin{equation}
\hat{ q} \Psi_{n} (q) = n q_e \Psi_{n} (q)  \nonumber .
\end{equation}

The Schr\"odinger equation, in the charge representation, becomes a difference equation, 
\begin{equation}
-\frac{\hbar^2}{2 L q_e^2} \left[ \Psi(q + q_e) + \Psi(q - q_e) - 2 \Psi(q) \right] +  V(q) \Psi(q) = E \Psi(q), \nonumber 
\end{equation}
corresponding to a discrete hamiltonian $ \hat{H} = \hat{H_0} + V(q)$. One may define the discrete charge shift
operators $\hat Q$ and $\hat{ Q}^{\dag}$, such that

\begin{eqnarray}
\hat Q ~\Psi(q) &=& \Psi (q + q_e) \\
\hat Q^{\dagger} \Psi(q) &=& \Psi(q-q_e),
\end{eqnarray}
and satisfy simple commutation relationships,

\begin{equation}
\left[ \hat{q}, \Qop \right] = - \qe \Qop ,~~~\left[  \hat{q}, \Qdag  \right] = \qe \Qdag ,~~~
\Qdag \Qop =  \Qop \Qdag = 1.
\end{equation}

Using these operators, one may write the hamiltonian operator, and also define a discrete flux operator 
(momentum) $\hat{\Phi }= (\hbar/2 i \qe) ( \Qop - \Qdag ) $. In the limit $q_e \to 0$, the operator 
$\hat \Phi \to -i \hbar \partial /\partial q$, but for finite $q_e$, $\hat \Phi$ satisfies modified 
commutation relationships with $\hat q$ and $\hat H_0$,
\begin{eqnarray}
\hat{H_0} & = &-\frac{\hbar^2}{2 L \qe^2} ( \Qop + \Qdag - 2) \nonumber \\
\left[ \hat{q}, \hat{\Phi} \right] &=& i \hbar ( 1 + \frac{\qe^2}{\hbar^2} \hat{H_0} ) ,~~~
\left[ \hat{H_0}, \hat{\Phi} \right] = 0 \nonumber ,~~~
\left[ \hat{H_0}, \hat{q} \right] = i \hbar \hat{\Phi} \nonumber
\end{eqnarray}

The operator $\hat \Phi$ posesses "plane waves" eigenstates, in the charge representation, 
these are $\Psi(q) = \exp(i\phi q /\hbar)$, with eigenvalue $\lambda = \sin(q_e \phi/\hbar)/(q_e/\hbar)$, 
in which $\phi$ is a quantum number with continuum eigenvalues, which may be called the "pseudo-flux". With
this one may define the "pseudo-flux" representation, in which one deals with "wavefunctions" $\Psi(\phi)$,
the charge operator $\hat q = i \hbar \partial /\partial \phi$, and the hamiltonian operator becomes

\begin{equation}
\hat H = -\frac{\hbar^2}{2C}\frac{d^2\Phi}{d\phi^2}+ \frac{2 \hbar^2}{L q_e^2} \sin^2(q_e \phi/2\hbar).
\end{equation}
Observe that the physical flux is the operator $\hat \Phi = -i \hbar \sin(q_e \hat \phi /\hbar)$, not $\hat \phi = \phi$, 
as in the "true" flux (momentum)representation. Note also that the physical current operator is obtained
from the canonically invariant definition $\hat I = -i [ \hat H, \hat q] /\hbar$, which gives $\hat I = \hat \Phi /L$, 
which is why we consider $\hat \Phi$ to be the physical flux.

\section{Previous applications}

We have applied this model to a few quantum systems, being able to reproduce some of the main results
obtained. In a previous work, Flores~\cite{FLORES} has proposed a quantum hamiltonian for a transmission line
with charge discreteness.  The periodic line is composed of cells containing capacity-coupled inductances, 
and charge discreteness is included as it has been described previously. One finds that charge discreteness 
implies that the charge operators at each cell satisfy nonlinear equations of motion; this difficulty has allowed 
so far only simplified, tight-binding type, calculations for the energy spectrum.

For the record, let us write the hamiltonian for the transmission line, obtained from the classical 
hamiltonian~\cite{FLORES} for $N$ cells,

\begin{equation}
\hat H = \sum_{l} \frac{2\hbar^2}{ L q_e^2} \sin^2( \frac{ q_e \phi_l}{2 \hbar} ) +
\frac{\hbar^2}{2 C} ( \hat q_{l+1} - \hat q_l )^2
\end{equation}

The Heisenberg equations of motion become

\begin{eqnarray}
\frac{d}{dt} \hat q_l &=& \frac{\hbar}{L q_e}\sin(q_e \phi_l/\hbar) \\
\frac{d}{dt} \hat \phi_l &=& \frac{1}{C} ( \hat q_{l+1} + \hat q_{l-1} - 2 \hat q_l) .
\end{eqnarray}

Recently, Flores et. al~\cite{FLORES-BOLOGNA} have found a nonlinear traveling wave solution to the equations
of motion above, in the long wavelenght limit, i.e., by replacing the differences by partial derivatives
in the above equations. The simplest solutions found look like sharp boundaries that travel with
a finite, constant, speed. 
 
Other system that has been studied is the open electron resonator (see Duncan et.al ~\cite{DUNCAN}). It has 
attracted considerable attention due to its remarkable behaviour (conductance oscillations). It has been 
explained by detailed theories based on the behaviour of electrons at the top of the Fermi sea, and also 
semiclassically~\cite{DUNCAN}. We studied this system using the simple quantum electrical circuit 
approach~\cite{LI-CHEN,YOU-LI,FLORES}, using a simple capacitor-like model of the system, being able to 
theoretically reproduce some features of the observed conductance oscillations~\cite{UTRERAS-FLORES}.

\section{Semiclassical study of $LC$ circuit}

The electrical engineer has learned to love the simplified description provided by the circuit 
description of a system, when compared with the more complete, but also more complex field description.
Nature is quantum, we say; however, we describe the behaviour of electrons modern circuits using the 
same basic laws (Kirchhoff) as in a classical circuit. There are difficulties on sight fot this state of
things, since we are now probing nature at very low temperatures, with very pure materials, very tiny 
currents and strong magnetic fields. Many examples show that things are about to change for the engineer,
we have seen flux quantization on superconductors, conductance oscillations, quantum hall effects (integer
and fractional), persistent currents and so on. 

It would be very useful to find out to what extent a circuit-like description could be of use for the very 
small electronic circuits of tomorrow, and what may be retained from it. We have been pursuing a description
that has the value of simplicity, some questions may be answered, but others perhaps not; for exaample, one 
area in which the "quantum $LC$ circuit" may give valid results is in the calculation of energy spectra. 
We have done this for the open resonator; but one may question that the calculation was not "simple enough", since
it still required to solve the Schr\"odinger equation. 

Now, we propose to go one step further in our simplification, by proposing to use a "semiclassical" approach 
(not in any strict sense, as it will be seen). We start from our hamiltonian, for the $LC$ circuit, with 
quantized electric charge 

$$
H = \frac{q^2}{2C} + \frac{2\hbar^2}{L q_e^2} \sin^2(q_e \phi/2\hbar).
$$

Consider this as a {\em "classical"} hamiltonian, with conjugate variables $\phi$ and $q$, 
we may write the corresponding hamilitonian equations (coincide with the Heisenberg eqs. too )

\begin{eqnarray}
\frac{\partial H}{\partial q} &=&  \frac{q}{C} = -\dot \phi \\
\frac{\partial H}{\partial \phi} &=& \dot q = \frac{\hbar}{L q_e} \sin(q_e \phi/\hbar) 
= \frac{\phi_0}{L} \sin( \phi/\phi_0).
\end{eqnarray}

To solve these classical equations ($q$ and $\phi$ are $c$-numbers), define now the dimensionless 
variable $\theta = \phi/\phi_0$, with $\phi_0 = \hbar/q_e$, $\Omega_0 = 1/\sqrt{LC}$, we obtain the 
same equation than {\em simple pendulum}, a system that has been quantized by several authors.
\begin{equation}
\ddot \theta = -\Omega_0^2 \sin(\theta),
\end{equation}

One simple way to obtain the approximate, quantum energy eigenstates, is provided by the old quantum
theory, as in the Bohr-Sommerfeld quantization rules, i.e., $\oint Q d\phi = (n +\gamma) h$ ($\gamma$ is 
a constant, we take $\gamma = 1/2$). To do this, let us express all in terms of the variable $\theta$ 
($E$ is the energy),

\begin{eqnarray}
Q &=& C \dot \phi = C \phi_0 \dot \theta \\
\phi &=& \phi_0 \theta \\
E &=& C\phi_0^2  \left( \frac{\dot \theta^2}{2} + 2\Omega_0^2 \sin^2(\theta/2 ) \right)
\end{eqnarray}
Assume that the maximum amplitude $\theta = \theta_0$, then the energy becomes 
$E =  2 C\phi_0^2 \Omega_0^2 \sin^2(\theta_0/2 )$, then we may obtain the "velocity" as a function of 
$\theta$, $\dot \theta = \dot \theta (\theta)$ 

\begin{eqnarray*}
\frac{ \dot \theta^2 }{2} &=& 2 \Omega_0^2 \left[ \sin^2(\theta_0/2) -  \sin^2(\theta/2) \right] \\
\dot \theta &=& \pm 2 \Omega_0 \sqrt{ \sin^2(\theta_0/2) -  \sin^2(\theta/2) }.
\end{eqnarray*}

From the above equations one may obtain $\theta (t)$, from the incomplete elliptical integral of the
first kind

\begin{equation}
\Omega_0 t = \int \frac{d\theta}{2 \sqrt{ \sin^2(\theta_0/2) -  \sin^2(\theta/2) }}.
\end{equation}

\subsection{Approximate energy states}

To find the (quantum) energy eigenstates, we follow a Bohr-Sommerfeld-like procedure, 
for which we need the phase -space integral $I$,
\begin{equation}
I = \oint Q d\phi = C \phi_0^2 \oint  \dot \theta (\theta) d\theta = 2 C \Omega_0 \phi_0^2 \oint \dot \theta d\theta.
\end{equation}

Define now $J =\oint \dot \theta d\theta$, then $I = 2 C \Omega_0 \phi_0^2 J$, and 

\begin{equation}
J = 4 \int_0^{\theta_0} \sqrt{ \sin^2(\theta_0/2) -  \sin^2(\theta/2)} d\theta,
\end{equation}
is an elliptical integral of the second kind, which may be expressed in the standard form by the
transformation $\sin (\xi) = \sin(\theta/2)/\sin(\theta_0/2)$,

\begin{eqnarray}
J &=& 8 \sin^2(\theta_0/2) \int_0^{\pi/2} \frac{ \cos^2(\xi)~d\xi }{\cos(\theta/2)} \\
&=& 8 \sin^2(\theta_0/2) \int_0^{\pi/2} \frac{ \cos^2(\xi)~d\xi }{\sqrt{1 - \sin^2(\theta_0/2) \cos^2(\xi)} }.
\end{eqnarray}

The complete elliptic functions of the first and second kind are defined, respectively, 
as (see Gradshteyn and Ryzhik, sec. 8.1~\cite{GRADSHTEYN})

\begin{eqnarray}
F(\phi,k)& =& \int_{0}^{\phi} \frac{d\alpha}{ \sqrt{1 - k^2 \sin^2(\alpha)} } \\
E(\phi,k) &=& \int_{0}^{\phi} \sqrt{1- k^2 \sin^2(\alpha) }~d\alpha
\end{eqnarray}

In our case, we identify $k = \sin(\theta_0/2)$ and $k^{\prime} = \sqrt{ 1 - k^2} = \cos(\theta_0/2)$. The integral 
that we seek is given in the tables ~\cite{GRADSHTEYN}, page 162 (2.584, \#6)

\begin{equation}
\int \frac{\cos^2(x) ~dx}{\sqrt{1 - k^2 \sin^2(x)}} = \frac{1}{k^2} E(x,k) - \frac{k^{\prime 2} }{k^2} F(x,k).
\end{equation}

Therefore, we obtain $J$ in terms of the complete elliptic integrals, $K(k) = F(\pi/2,k)$ and $ E (k) = E(\pi/2,k)$,
\begin{equation}
J = 8 \left( E(k )) - (1 - k^2) K(k )\right).
\end{equation}

The complete elliptical integrals may be expanded in series, as follows (Gradshteyn~\cite{GRADSHTEYN}, pp. 905)

\begin{eqnarray}
K &=& \frac{\pi}{2} \left\{ 1 + (\frac{1}{2})^2 k^2 + ( \frac{3}{8})^2 k^4 + \dots \right\} \\
E & =& \frac{\pi}{2} \left\{ 1 - \frac{1}{2^2} k^2 - \frac{3}{2^2 \cdot 4^2} k^4 + \dots \right\}
\end{eqnarray}

With this, $J$ becomes, to fourth order in $k$,

$$
J = 2\pi k^2 ( 1 - \frac{5}{32}k^2 + \dots ),
$$
and, assuming $I = (n + \gamma) h$, ($h=$ Plank's constant), the quantization rules of the
old quantum theory, we get
\begin{equation}
I = \frac{4 \pi \phi_0^2}{L \Omega_0} k^2 ( 1 - \frac{5}{32}k^2 + \dots ) = (n + \gamma) h
\end{equation}

Assuming $k << 1$ we have, to first order,

$$
(k_n^{(0)})^2 = \frac{(n+ \gamma) \hbar L \Omega_0}{2 \phi_0^2},
$$
the corrected $k_n$ becomes
$$
k_n^2 = (k_n^{(0)})^2 ( 1 - \frac{5}{32}(k_n^{(0)})^2 ) .
$$
The energy levels become (we use $\gamma = 1/2$), up to second order 

$$
E_n = C \phi_0^2 \times 2 \Omega_0^2 k_n^2  \approx \hbar \Omega_0 (n + \gamma) 
\left( 1  - \frac{5}{32}\frac{(n+ \gamma) \hbar L \Omega_0}{2 \phi_0^2} \right).
$$

Another interesting case arises when one has $k$ close to 1, then, the relevant expansion parameter
is $k^{\prime} = \sqrt{1 - k^2}$.

\subsection{ $C$-design case}

The so-called $C$-design case corresponds to the 'large capacity' case, in which the electrostatic part
dominates. We have seen that the 'velocity' $\dot \phi$ may be written as

$$
\dot \phi = \pm \sqrt{\frac{2 E}{C}} \left[ 1 - \frac{2 \phi_0^2}{L E} \sin^2(\frac{\phi}{2\phi_0}) \right]^{1/2}.
$$

Consider the case in which the energy $E >> \phi_0^2/2L$, i.e., in which the electrostatic energy dominates,
the so-called $C$-design, then one may expand in a series, keeping only the first two terms ( let $Q = C\dot \phi$),

$$
Q \approx \pm \sqrt{2 C E} \left[ 1 - \frac{\phi_0^2}{2 L E} \left( 1 - \cos(\frac{\phi}{\phi_0}) \right) \right],
$$
and, as before, let $\Delta \phi = 2\pi \phi_0$, and compute the action variable $I$,

\begin{equation}
I = \int_0^{\Delta \phi} Q ~d\phi \approx  \sqrt{2 C E} \Delta \phi \left[ 1 - \frac{\phi_0^2}{2 L E}  \right],
\end{equation}
from this, letting $I = n h$, we obtain the energies
\begin{equation}
E_n \approx \frac{(n q_e)^2}{2 C} \left( 1 + \frac{ \phi_0^2/L}{ (n q_e)^2/2 C} \right) = \frac{(n q_e)^2}{2 C} + \frac{\phi_0^2}{L}.
\end{equation}

\section{Final remarks}

We apply here a method that we call {\em semiclassical}, to solve the equations of the discrete-charge quantum 
circuit; the reason for this name is due to the obvious similarity with the WKB approximation, however, we note here 
that the very existence of the {\em quantum circuit} idea has no classical analog. Instead, the classical analog
would require the existence of a flux quantum;  in other words, the 'classical analog' of the quantum circuit
is a classical object with quantized flux. Starting from there, one may apply the WKB method, with its
semiclassical (Bohr-Sommerfeld) rules, for the quantization. This has been done here, noting the strong 
similarity of our system and the quantized simple pendulum.

\ack{The authors aknowledge the finantial support provided by FONDECYT Grant \# 1040311, and DIDUACH
Grant \# S-2004-43. C.A. Utreras-D\'{i}az also acknowledges support from Universidad Austral de Chile (DID Grant \#
S-2004-43). Useful discussions originally were carried-out with V. Bellani,
(Pavia University), M. Bologna (Universidad de Tarapac\'a), and A. Zerwekh (Universidad Austral de Chile).

\end{document}